\documentclass[doublecol]{epl2} 
\usepackage{graphicx}
\usepackage{subfigure}
\usepackage{epstopdf}
\usepackage{amsmath} 
\usepackage{subfigure}

\title{Synchronization in a stochastic Hebbian network of phase oscillators}

\author{A. Isakov \inst{1},~ L. Mahadevan \inst{1,2}}

\institute{                    
\inst{1} Department of Physics, Harvard University, Cambridge, MA. \\
\inst{2} School of Engineering and Applied Sciences, Harvard University, Cambridge, MA.}


\abstract{ We investigate a generalized Kuramoto phase-oscillator model with Hebb-like couplings that evolve according to a stochastic differential equation on various topologies. Numerical simulations show that even with identical oscillators, there is a regime in the nearest-neighbor coupling topologies and a complex network topology where oscillators move between an in phase and anti-phase state. Phase diagrams show the transition probabilities as a function of the noise strength and rate of evolution of network coupling. A minimal theoretical model allows us to understand these transitions.}

\begin{document}

\maketitle

\section{Introduction}
\indent A fundamental question in studying self-organizing phenomena is determining what broad classes of conditions lead to which types of possible states of synchronization. From a mathematical point of view \cite{ref:1,ref:2}, symmetry considerations can provide some guidelines for possible equilibrium states, with certain limitations on the allowable types of systems. Perhaps one of the simplest general models of synchronization is the case of coupled phase oscillators \cite{ref:5}. The framework of the eponymous model was analyzed and popularized by Y. Kuramoto \cite{ref:10} and has since seen extensive study in physics, biology, game theory, and other disciplines as a stepping stone towards understanding more complex systems \cite{ref:12,ref:14,ref:16, ref:18}. In the simplest case, the $N$ oscillators are described by
\begin{equation} \label{eq.1} \dot{\phi}_{i} = \omega_{i} - \frac{K}{N}\sum_{j \in S_{i}} \sin(\phi_{i}-\phi_{j}), ~~ i = 1,\ldots,N, \end{equation}
where $S_{i}$ is the set of neighbors of oscillator $i$, not including itself. Here $\phi_{i}(t)$ is the phase of oscillator $i$ at time $t$ and $\omega_{i}$ is its natural frequency. The coupling $K$ is usually assumed to be constant and positive, leading to attractive interactions: slower oscillators are sped up by their neighbors and faster oscillators slow their neighbors down. \\
\indent This model acts as a springboard to a set of natural generalizations such as changing the connectivity topology $S$, drawing $\omega_{i}$ from different distributions, adding noise terms to the equation, and considering more general coupling functions \cite{ref:20}. \\
\indent Recently, biological and social network motivations have provided the need to consider activity-dependent interactions. It is now known that changes in connections between neurons are related to the relative times between firing or the synchronization between them \cite{ref:30, ref:31}. Even in the classical examples of synchronization in clapping audiences \cite{ref:32} and circadian rhythms \cite{ref:34} one can discern that the coupling changes as a function of the current synchronization of the system. In order to provide a more accurate model for real systems, we need to consider the case when the coupling strength obeys its own dynamics on a time scale which is slow compared to the phase dynamics of the oscillators \cite{ref:40,ref:42,ref:44,ref:46}, in the spirit of a Hebbian network \cite{ref:Hebb}. In neural networks, this corresponds to having neurons which fire together to wire together slowly. In social phenomena, this may correspond to the ties between people changing based on the behavior of the people themselves \cite{ref:48}. \\
\section{Model}
\begin{figure*}[ht]
\includegraphics[width=2\columnwidth]{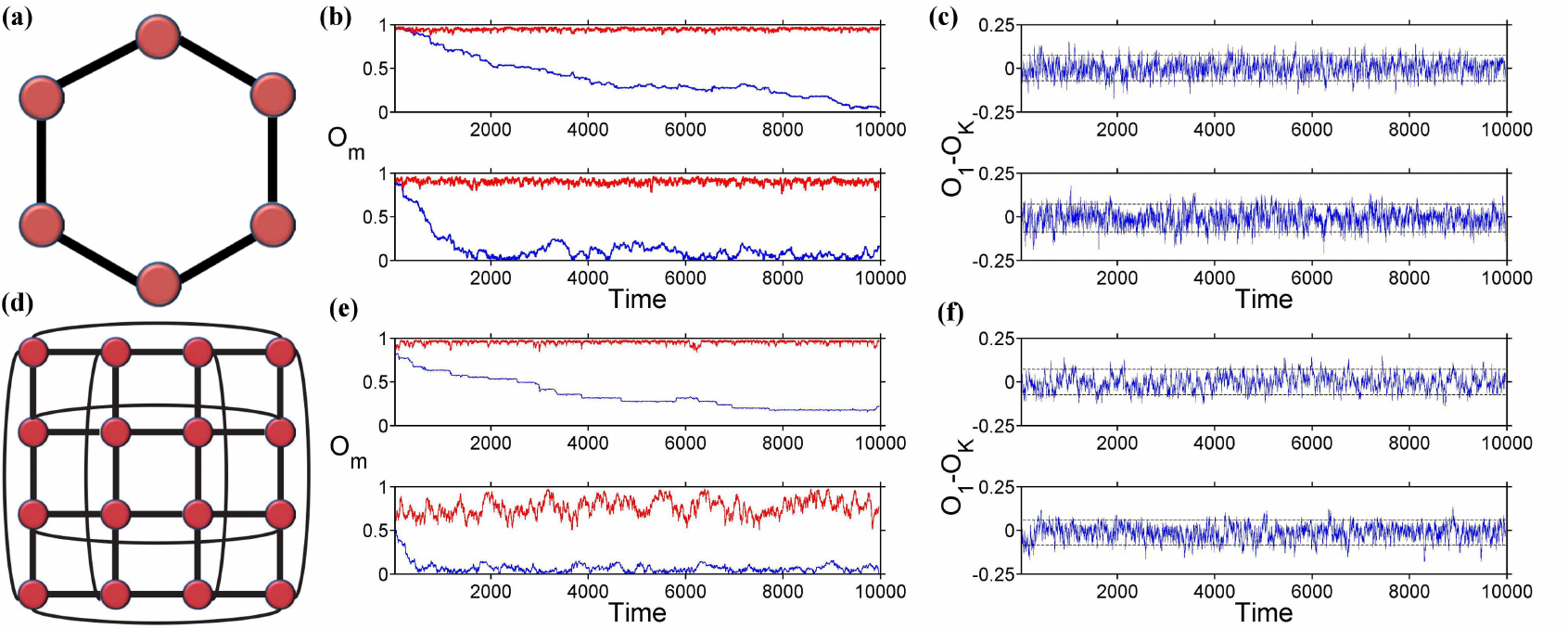}
\caption{{\bf Topology and dynamics of oscillators with nearest neighbor coupling in 1D and 2D.} All data in figures is from numerically solving eqs. (\ref{eq.2},\ref{eq.3}). Order parameters $O_{m=1,2},O_{K}$ are as defined in eqs. (\ref{eq.4},\ref{eq.5}). Parameters: $N = 100$, $\sigma = 0.2,~0.25$ (top to bottom). Top row is for ring topology, bottom row is for 2D nearest neighbor topology. Data starts at $t = 30$ (300 time-steps) to remove transients. Panels (b, e) show that the system can maintain overall synchrony (high $O_{2}$) even at small $\epsilon$ and large $\sigma$. $O_{1}$ tends to move down in a slow step-wise fashion for lower $\sigma$ at a fixed $\epsilon$, indicating the existence of relatively rare ``discrete'' events. At higher $\sigma$, $O_{1}$ drops very rapidly, so the oscillators quickly become arranged randomly in relative phase/anti-phase. Panels (c, f) show that $O_{K}$ tracks well with $O_{1}$ - the difference looks randomly distributed around $0$ and is within tight bounds. They are good mutual predictors even for relatively small $\epsilon$ and large $\sigma$. (a) Illustration of ring topology - nodes (circles) with connecting lines are coupled. (b) $O_{1}$ (blue) and $O_{2}$ (red). $\epsilon = 0.1$. (c) $O_{1}-O_{K}$ (blue). $\epsilon = 0.1$. The dashed lines encompass $90\%$ of the data points on either side of $0$. (d) Illustration of 2D nearest neighbor topology with periodic boundary conditions. (e) $O_{1}$ (blue) and $O_{2}$ (red). $\epsilon = 0.05$. (f) $O_{1}-O_{K}$ (blue). $\epsilon = 0.05$. The dashed lines delineate $90\%$ of the data points on either side of $0$.}
\label{ref:fig1}
\end{figure*}

\indent We start with a modified Kuramoto model where the slowly evolving coupling obeys the simplest expected symmetry laws and is affected by a noisy environment, more closely mimicking a real biological or social system. For simplicity, we assume identical oscillators ($\omega_{i = 1,\ldots,N} = \omega$) and rescale by $\omega$ such that the oscillators have a natural frequency of $1$. The set of equations for the system is \cite{ref:40}
\begin{equation} \label{eq.2} \dot{\phi}_{i} = 1 - \alpha\sum_{j \in S_{i}} K_{ij}\sin(\phi_{i}-\phi_{j}) + \sigma_{1}\zeta_{1;i}, ~~ i = 1,\ldots, N\end{equation}
\begin{equation} \label{eq.3} \dot{K}_{ij} = \left\{
\begin{array}{l l}      
    \epsilon(-K_{ij}+\cos(\phi_{i}-\phi_{j}))+\sigma\zeta_{2;i,j}, & j \in S_{i}, \\
    0 & \text{otherwise}.
\end{array}\right.
\end{equation}
To ensure that each neighbor contributes a fraction proportional to the number of neighbors to the frequency of oscillator $i$, we let $\alpha=\frac{1}{|S_{i}|+1}$; here the form of the denominator is chosen to allow the oscillator to ``affect itself''. The long time scale $\tau \sim \epsilon^{-1} \gg 1$ describes the slow evolution of the network coupling. We note that the symmetry of coupling with $K_{ij} = K_{ji}$ is preserved by the functional form chosen. The noise  $\zeta$  is assumed to be Gaussian with $\langle \zeta \rangle = 0,~ \langle \zeta_{i}(t)\zeta_{j}(t')\rangle = \delta(t-t')\delta_{ij}$, and standard deviation $\sigma_{1},~\sigma$. We take $\sigma_{1} \ll 1$  to prevent the system from locking when $\phi_{i} - \phi_{j} = 0$ or $\pi$. Thus, the model is characterized by the two parameters $\epsilon, \sigma$. \\
\indent In order to complete the model description we must specify a coupling topology.  We start by considering both a ring topology (fig. \ref{ref:fig1}(a)) and a 2-dimensional nearest neighbor model with periodic boundary conditions (fig. \ref{ref:fig1}(d)). Later, we will discuss the case of a modified preferential attachment model \cite{ref:49} whose mean mimics the properties of the 2-dimensional lattice. For the ring, labeling oscillators from $0$ to $N-1$, the non-vanishing coupling variables  are $K_{i,\text{mod}(i \pm 1,N)}$ for $i = 0,\ldots,N-1$. This labeling carries over analogously to the two-dimensional case by adding a second index to the oscillators to arrange them ``on a grid''. From now on all subscripts are modulo $N$, so oscillator $i$ connects to oscillators $i \pm 1$ for the ring and oscillator $(i,j)$ connects to $(i\pm 1, j)$ and $(i,j\pm 1)$ in the 2D case. Thus, we have $|S_{i}| = 2,~5$ and $\alpha = \frac{1}{3},~\frac{1}{5}$ for the respective topologies. \\
\section{Discussion}
\indent We explore the system behavior using  by solving the stochastic eqs. (\ref{eq.2},\ref{eq.3})  iteratively in MATLAB using the Euler-Maruyama method with scaled time-step $\Delta t = 0.1$ for $10^{5}$ steps. We start with initial conditions $K_{ij} = 0$ and $\phi_{i} = 0$ for all $i,j$; random initial conditions did not change any qualitative results. For all simulations we use $N = 100$ oscillators, thus approaching the thermodynamic limit. \\
\indent We find that the appropriate oscillator order parameters for the ring topology are
\begin{equation} \label{eq.4} O_{m} = \frac{1}{N}\left|\sum_{j=1}^{N}e^{im(\phi_{j}-\phi_{j-1})}\right|,~~ m = 1,2, \end{equation}
where $i = \sqrt{-1}$. Longer-range order parameters were also tried but were not as successful at capturing the results. Values of $O_{2}$ close to unity imply that all oscillators are coherent, with $\phi_j - \phi_{j-1}=\{0, \pi\}$, while values of $O_{1}$ close to unity imply that most oscillators are in phase, with $\phi_j - \phi_{j-1}=0$. For the coupling coefficients, we found that the order parameter that captures the link to the dynamics of the system of oscillators (in particular, to $O_{1}$) is the simple expression
\begin{equation}\label{eq.5} O_{K} = \frac{1}{N}\left|\sum_{i=0}^{N-1} K_{i,i-1}\right|. \end{equation}
Figure \ref{ref:fig1}(b) shows three representative plots of the order parameters $O_{m=1,2}$ using $\epsilon = 0.1$ and $\sigma = 0.20,~0.25$ for the top and bottom panels, respectively, starting at $300$ time-steps to remove the transient from the deterministic initial conditions. As $\sigma$ increases, $O_{2}$ remains near $1$ but with a larger spread. Thus, the effect of noise on coupling does not throw the system into disarray - there is still a sense of synchronization. However, as the noise $\sigma$ increases, the coupling is more disordered and the oscillators can move from in phase to anti-phase relations with their neighbors. Then, the oscillators switch between two possible discrete phases (a form of synchrony) but move between them incoherently. Of course, for very large $\sigma$ (outside of the range shown), there is an uninteresting incoherent regime characterized by $O_{2} $ decreasing significantly from unity. \\
\indent Figure \ref{ref:fig1}(c) shows the difference $|O_{1} - O_{K}|$ for the same two sets of parameters starting at $300$ time-steps. We see that they stay within a relatively tight band - the dashed lines delineate where $90\%$ of the data lines on either side. The fact that the lines are highly symmetric about $0$ and have magnitude near $\pm 0.1$ even in the high $\sigma$ case suggests that $O_{K}$ tracks $O_{1}$ very closely. \\
\indent The order parameters generalize directly to the 2D case by adding a second index and changing the constant in front to the total number of couplings (number of unique connected pairs). Figure \ref{ref:fig1}(e, f) show $O_{m=1,2}$ and $O_{1}-O_{K}$ in the 2D topology for $\sigma = 0.20,~0.25$ for the top and bottom panels, respectively. A similar pattern as in the ring topology emerges, where oscillators have a high relatively high overall synchrony and tend towards arranging in random phase/anti-phase relations in time. Again, $O_{K}$ follows $O_{1}$ closely, as seen by tight bounds of the dashed black lines representing the cutoff for $90\%$ of the positive and negative points. In fact, even though we consider a lower $\epsilon = 0.05$ (as compared to $\epsilon = 0.10$ for the ring) and the same $\sigma$, the lines form a tighter band in the 2D nearest neighbor topology. This opens up the intriguing possibility of reasoning about the order parameter of the coupling strength, which is in general hard to observe, based on the more easily observable synchronization of the oscillators, especially in more realistic scenarios where a 2D structure provides a better model. \\

\begin{figure*}[ht]
\includegraphics[width=\columnwidth]{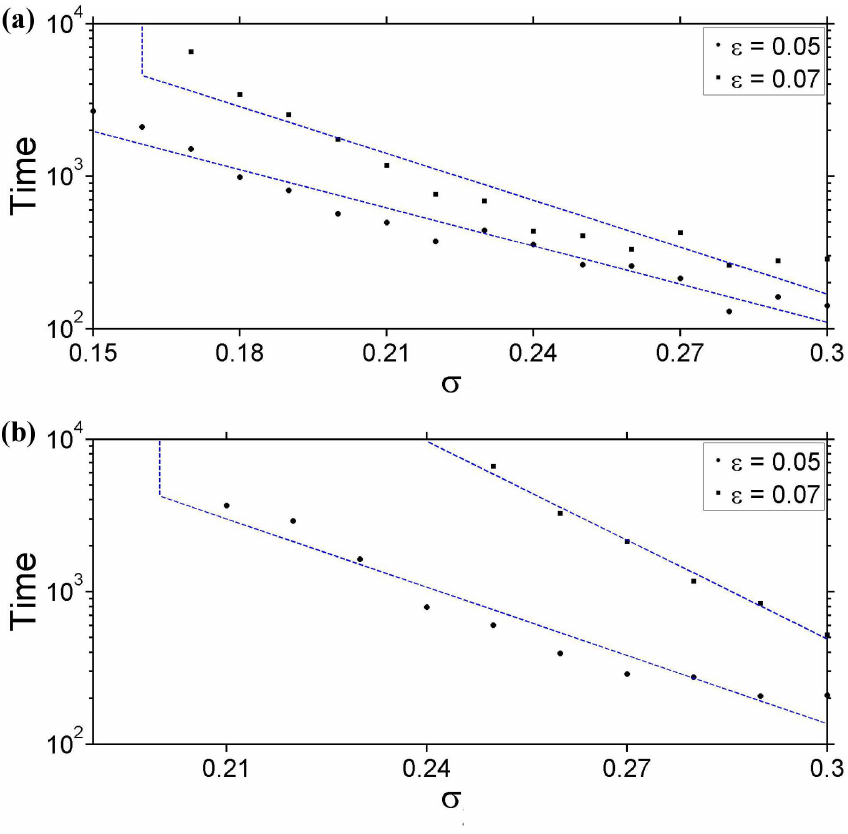}
\caption{{\bf Time for desynchronization in 1D and 2D topologies as a function of the noise strength $\sigma$ for different values of the rate of change of coupling $\epsilon$.} Time required for $O_{1}$ to reach $10\%$ of its starting value of $1$ as a function of $\sigma$ for various values of $\epsilon$ (logarithmic scale). The dashed vertical lines represent the point before which $O_{1}$ did not reach $0.1$ during the full simulation run-time. Non-vertical dashed lines are best fits. $N = 100, \epsilon = 0.05,~0.07$ (circle, square). The time to crash is close to exponential in $\sigma$, indicating the existence of distinct regimes. Topologies with more connections take longer to crash for the same parameter values. (a) Ring topology. (b) 2D nearest neighbor topology.}
\label{ref:fig2}
\end{figure*}

\indent To understand the temporal dynamics of synchronization, in fig. \ref{ref:fig2}(a, b) we plot the logarithm of the time required for $O_{1}$ to ``crash'' (defined as first achieving a threshold value $0.1$ of the initial value $O_{1}(0) = 1$) as a function of $\sigma$ for representative values of $\epsilon$ in the ring and 2D  topologies. All results are averaged over $10$ runs, and dashed lines correspond to the best fit. A value is not plotted if $O_{1}$ did not reach $0.1$ during the full simulation time.  This suggests that for small $\sigma$ oscillators switch between alignment and anti-alignment rarely, while for large $\sigma$ switching is common. Guided by this intuition, we explore the possibility of distinct synchronization regimes in the system as a function of the noise strength $\sigma$ and the rate of change of coupling $\epsilon$.
\begin{figure*}[ht]
\includegraphics[width=\columnwidth]{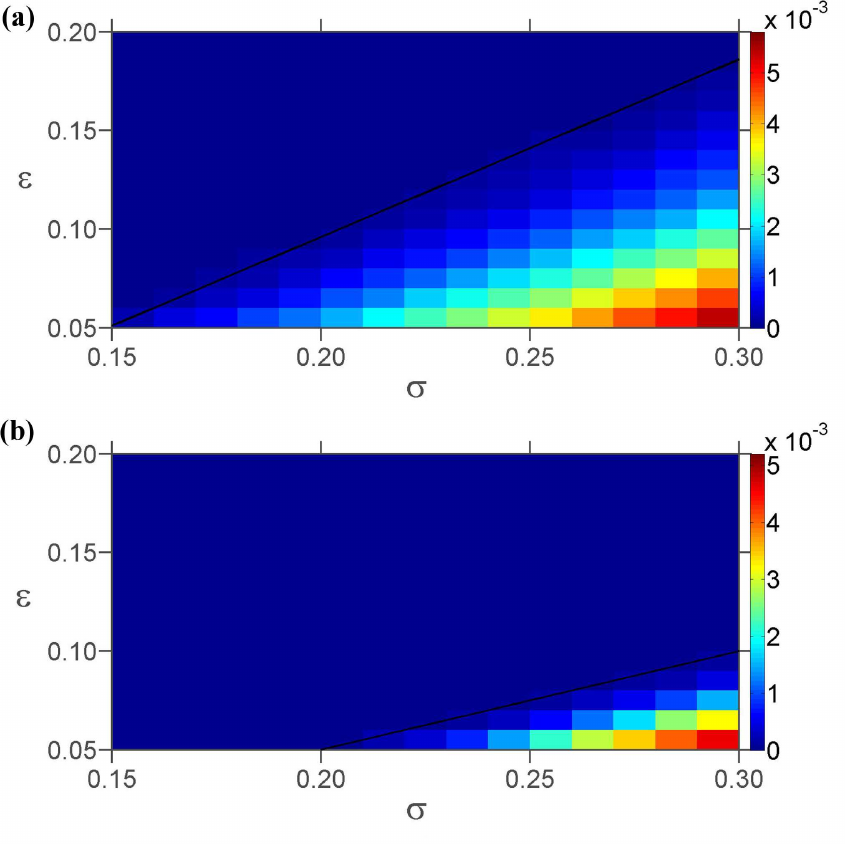}
\caption{{\bf Phase diagrams for synchronization/desynchronization in 1D and 2D nearest neighbor networks as a function of the noise signal strength $\sigma$ and rate of change of coupling $\epsilon$.} Color corresponds to the frequency of switching - blue is low, red is high. Values are per oscillator per unit time. Black lines correspond to the approximate division between the synchronized regime (no switching) and the dynamic (switching) regime. $N = 100$. Two distinct regimes are visible: a synchronous regime above the line and a ``flipping'' regime below the line. (a) Ring topology. (b) Nearest neighbor topology.}
\label{ref:fig3}
\end{figure*}
\\ \indent Figure \ref{ref:fig3}(a, b) shows phase diagrams in the $\epsilon-\sigma$ plane that show the how the system can operate in two distinct regimes split approximately along a straight line in the ring and 2D topologies, respectively. We define a flip to be an event where the difference of phases between two neighbors changes from $(\Delta \phi \text{ mod}~ 2\pi) \in \left[0,\frac{\pi}{3}\right]\cup\left[\frac{5\pi}{3},2\pi\right]$ to $(\Delta \phi \text{ mod}~ 2\pi) \in \left[\frac{2\pi}{3},\frac{4\pi}{3}\right]$ or vice versa. Contracting the bounds did not affect the results. Results were averaged over the number of pairs over $10$ runs. In the ``synchronous regime'' (above the line in fig. \ref{ref:fig3}), the oscillators achieve a steady state where after an initial fluctuation they choose to be in a phase relation of $0$ or $\pi$ with their neighbors, depending on the couplings, and freeze. However, for larger values of $\sigma$ for a fixed $\epsilon$, they dynamically switch between in phase and anti-phase relations with their neighbors. We call this the ``flipping regime'' (below the line). Aside from a fast transient, even in the flipping regime there is no ``incoherent'' motion - a particular oscillator $i$ simply changes from being closely synchronized with its neighbors to being anti-synchronized. Thus, the system maintains overall coherence while each oscillator can move between two states. For the ring topology, the line shown is $\epsilon = 0.9\sigma-0.085$ while in the 2D case the line is $\epsilon = 0.5\sigma-0.05$. We do not show the region corresponding to completely asynchronous, noise-dominated motion, although from fig. \ref{ref:fig1}(e) we can see that the system begins to approach that regime towards the bottom right of the phase diagram. It is interesting to note that the line in the 2D topology has a significantly smaller slope than in the 1D case. Indeed, in the limiting case of an all-to-all connection topology (not shown), no flips were observed on the time-scale of the simulation in this parameter regime. \\
\indent The oscillators' behavior can be intuitively seen as follows: since they are identical and have fast dynamics compared to the coupling, they quickly arrange themselves into a steady state. When the coupling between neighbors changes such that the anti-phase state becomes more stable than the in phase state (or vice versa), the relevant oscillators quickly rearrange themselves to the new state, resulting in a high $O_{2}$. However, as the relative coupling strengths between an oscillator's neighbors shift more quickly, there are more flips between the two states the oscillator wants to adopt relative to its neighbors, leading to a lower $O_{1}$. If $\sigma$ is too high, the effects of neighbors are dominated by noise and the system is thrown into incoherence. Further, if there are too many neighbors, in order to have flipping occur one needs to have the majority of oscillators flip at once (since otherwise the other majority will tend to prevent flipping). Thus, topologies with many  neighbors lack the flipping regime. In other words, the flipping regime occurs in a special intermediate region and can be encapsulated as follows: there need to be few enough neighbors that their individual effects matter and the coupling between oscillators has to have sufficient noise to promote flipping but not so much that oscillators consistently have only random interactions and fall completely out of coherence. \\
\indent These observations naturally lead to considering the case of topologies with structural randomness to better understand how spatial variation interacts with the temporally noisy dynamics, which is of paramount importance in neural systems \cite{ref:52}. We use a modified preferential attachment model \cite{ref:55, ref:56} for $N = 100$ oscillators with a tuning parameter $p$: with probability $p$, a new node attaches randomly to one of the previous nodes, and otherwise it attaches to a previous node $i$ with probability equal to $\frac{k_{i}}{\sum_{j=1}^{i-1}k_{j}}$, where $k_{i}$ denotes the degree of node $i$. Then, we add a ``backbone'' of a ring lattice to allow comparison with the 2D lattice. We consider each new node making $l = 1,~2,~3$ attachments to prior nodes; the case $l = 1$ corresponds to a mean degree of approximately $4$, as in the lattice case, while the case $l = 2$ corresponds to a minimum of $4$ neighbors (up to removal of duplicate links when adding the ring lattice backbone, which was negligible) with a mean degree of approximately $6$. The degree distribution when $l = 3$ has a mean of approximately $8$.
\\ \indent The phase diagrams for flipping times in $p$-$\sigma$ space are shown in fig. \ref{ref:fig4}, considering flips only between neighbors along the ring, with $\epsilon = 0.05$. While the results are seen to be independent of $p$, they are strongly dependent on $l$. For $l = 1$, the flipping rates are generally of the same order of magnitude as in the $2D$ lattice, while for $l = 2$ flipping begins at $\sigma = 0.27$ and is of significantly smaller magnitude compared to the same $\sigma$ when $l = 1$. The flipping regime is almost extinguished when $l = 3$ - flipping rates are on the order of $10^{-4}$. Thus, although spatial variability itself may not be usable as a valid method for the fine control of temporal dynamics, the synchronization of this system is highly robust to the underlying network topology and depends on the mean of the underlying degree distribution. This result is in line with the previous analysis where we found that the synchronization effects are only a function of local connectivity. Thus, even in the case of high spatial variance where some nodes are highly connected (and hence are not expected to flip often), there are so few of these nodes compared to those of low degree (below the average, hence more likely to flip) that their effect on the macroscopic system behavior is largely negligible. However, increasing the minimum degree of all nodes makes all nodes less likely to flip, impacting the macroscopic behavior. This robustness result is interesting because it helps us understand the wide-spread appearance of synchronization in real-life networks, which can exhibit any number of interesting topologies that also have slowly varying coupling strengths. \\
\begin{figure*}[htbp]
\includegraphics[width=1.8\columnwidth]{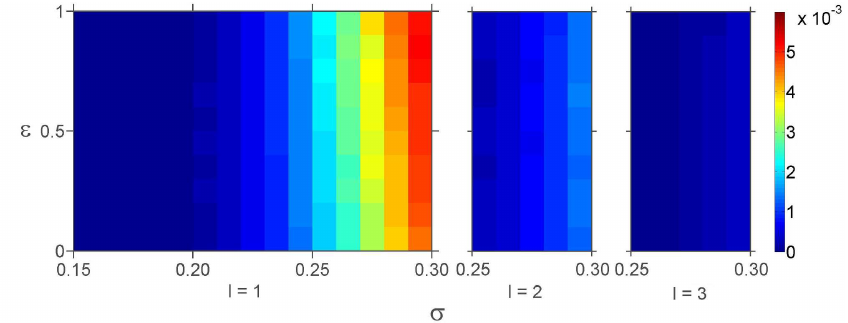}
\caption{{\bf Phase diagram for synchronization/desynchronization in a random network with preferential attachment and different values of mean degree.} Phase diagrams showing the number of times oscillators switch from an in phase to an anti-phase state relative to their neighbors on the ring as a function of $\sigma, p$ for three different values of $l$. Values are per oscillator per unit time. $N = 100, \epsilon = 0.05$. The amount of flipping is found to be independent of $p$ but highly dependent on $l$. When $l = 1$ (same mean degree as 2D lattice), the amount of flipping is approximately a factor of $2$ less than in the 2D lattice. In the case where $l = 2$ (same minimum degree as the lattice, mean degree = $6$), flipping is largely extinguished, and at $l = 3$ there is no consistent flipping even at the highest value of $\sigma = 0.30$.}
\label{ref:fig4}
\end{figure*}
\indent To get an analytical perspective on the system dynamics as a function of $\epsilon$, we can provide a useful approximation for $O_{1}(t)$ based on the results of the numerical experiment. Since the order parameter for the coupling $O_{K}$ tracks $O_{1}$, we can estimate $O_{1}(t) \sim O_{K}(t) = \frac{1}{N}\sum_{i,j}K_{i,j}$, dropping absolute value signs from the definitions since they were only introduced into the order parameters for convenience. Summing eq. (\ref{eq.3}) over all neighbors $i,j$ yields
\begin{equation}\label{eq.6} \sum \dot{K}_{ij} = \epsilon \left(-\sum K_{ij} + \sum \cos(\phi_{i}-\phi_{j})\right) + \sum \sigma\zeta_{2;i,j}. \end{equation}
Since $\langle \zeta_{2;i,j}(t) \rangle = 0$, the last term vanishes in the large $N$ limit. Approximating $\text{sign}(\cos(\phi_{i}-\phi_{j}))$ by $\text{sign}(K_{ij})$ because synchronization between neighbors occurs on a faster time-scale than the change in coupling we obtain
\begin{equation} \label{eq.7} \frac{d}{dt} \sum K_{ij} = -\epsilon\sum K_{ij} + \epsilon\sum \text{sign}(K_{ij}). \end{equation}
In the no flipping regime, the two terms on the right hand side cancel as each $K_{ij}$ settles into an average steady state of $\pm 1$. In the flipping regime, flips occur relatively rarely so the last term on the right hand side can be approximated as $\epsilon \beta$, where $\beta$ is approximately constant and bounded by the number of couplings (number of unique oscillator pairs). The solution to this equation is $\sum K_{ij} = Ce^{-\epsilon t} + \beta$, where $C$ depends on initial conditions. Indeed, the bottom panels of fig. \ref{ref:fig1}(b, e) show an approximately exponential decline in $O_{1}$ towards a state where there is overall synchrony in the system but each set of neighbors may be aligned or anti-aligned. \\
\section{Future Work}
\indent An interesting direction for future work is applying the model to experimentally observed topologies. It would be promising to look at the correspondence between the order parameters for the oscillators and the coupling strength with a view to reasoning about coupling strength from direct observations of e.g. neural networks. A local tie between coupling and the phase/anti-phase relation in the ring and 2D nearest neighbor topology is given by $\text{sign}(K_{i,i-1})$. In a typical run, this measure correctly predicts the synchronization relation between two neighbors ($\text{sign}(K_{i,i-1}) = \pm 1$ when $i,i-1$ are in phase and anti-phase, respectively) approximately $90\%$ of the time even for the extreme case $\epsilon = 0.05, \sigma = 0.30$ and even more accurately farther away from the bottom right corner of fig. \ref{ref:fig3}(a, b). Another promising line of research is to consider the proposed model in the context of chimera states \cite{ref:60,ref:62,ref:64}. Experimentally, it may be possible to test the phase diagram predictions with an electromechanical or laser model using feedback as the analogy to a strengthening or weakening coupling. \\
\indent Research conducted with Government support (A.I.) under FA9550-11-C-0028 and awarded by the Department of Defense, Air Force Office of Scientific Research, National Defense Science and Engineering Graduate Fellowship, 32 CFR 168a.

\end{document}